\documentclass[journal,onecolumn]{IEEEtran}

\usepackage[utf8x]{inputenc} % To use Unicode (e.g. Turkish) characters

\usepackage{amsmath, amsthm, amssymb}
\usepackage[bottom]{footmisc}
\usepackage{cite}
\usepackage{graphicx}
\usepackage{longtable}
\usepackage{url}
\usepackage{siunitx}
\DeclareSIUnit{\belmilliwatt}{Bm}
\DeclareSIUnit{\dBm}{\deci\belmilliwatt}

\graphicspath{{figures/}} % Graphics will be here
\usepackage{acronym} 

\usepackage{multirow}
\usepackage{subfigure}
\usepackage{algorithm}
\usepackage{algorithmic}
\ifCLASSINFOpdf
\else
\fi
\begin{document}
\title{More WiFi for Everyone: Increasing Spectral Efficiency in WiFi6 Networks using OBSS/PD Mechanism}

\author{Ali~Karakoç,
        Mehmet~Şükrü~Kuran,~\IEEEmembership{Member~IEEE},
        and~H.~Birkan~Yilmaz,~\IEEEmembership{Member,~IEEE}% <-this % stops a space
}

\maketitle

\begin{abstract}
This study aims to enhance spatial reuse by using the new features of IEEE 802.11ax WLANs. Since the wireless medium is a shared medium and there may be multiple basic service sets (BSS) in the same vicinity, BSSs may overlap, and interference occurs. In this situation, BSSs cannot transmit simultaneously due to the exposed node problem. The IEEE 802.11ax standard has a couple of mechanisms to resolve these spectral efficiency problems. One of the most effective mechanisms that address these problems is the overlapping BSS preamble detection (OBSS/PD) mechanism. OBSS/PD mechanism uses the color mechanism to distinguish OBSS signals. By using a signal threshold, the mechanism can ignore some of the signals, which cause interference. In this paper, we propose a rate-adaptive dynamic OBSS/PD threshold algorithm that tracks the changes in transmission rate and dynamically adjusts the threshold step by step considering the changes.
\end{abstract}

% \fi
\IEEEpeerreviewmaketitle

%%%%%%%%%%%%%%%%%%%%%%%%%%%%%%%%%%%%%%%%%%%%%%%%%%%%%%%%%%%
%%%%%%%%%%%%%%%%%%%%%%%%%%%%%%%%%%%%%%%%%%%%%%%%%%%%%%%%%%%
%%%%%%%%%%%%%%%%%%%%%%%%%%%%%%%%%%%%%%%%%%%%%%%%%%%%%%%%%%%
\section{Introduction}
\IEEEPARstart{T}{his} paper focuses on the WiFi standard IEEE 802.11ax and its features related to spatial reuse. Among the various features being proposed in WiFi6, one of them is the OBSS-PD mechanism. In dense environments, if there are overlapping BSSs using the same channels, the traffic belonging to these BSSs must share the channel and contend for access. Before performing the transmission, each node should check whether the medium is in an idle state or not.

In some cases, two nodes belonging to different BSSs cannot transmit simultaneously even if their transmissions will not cause collisions. At its core, the spatial reuse features of 802.11ax allow the STAs and APs to adjust their ranges dynamically to maximize the spectral efficiency. OBSS/PD mechanism is an extension to the legacy CSMA/CA mechanism. This paper proposes a rate-adaptive inter-BSS carrier elimination-based OBSS/PD threshold (RACEBOT) algorithm. Mainly, the RACEBOT algorithm is a received signal strength indicator (RSSI) based carrier sensing threshold algorithm, and it is designed to be used with typical rate selection algorithms. Since changing the OBSS/PD threshold also affects the transmission power, this operation is also related to rate selection algorithms. If the transmission power is not enough to transmit for a corresponding MCS level, the transmission fails, hence the RACEBOT algorithm also considers the transmission power for the rate selection algorithm stability and performance. Therefore, RACEBOT changes the OBSS/PD threshold with small steps. Besides, the RACEBOT algorithm looks at the count of each OBSS RSSI level, and it ignores the RSSI levels, which are below the count threshold. Therefore, OBSS/PD threshold can be chosen at lower levels, and due to the higher transmission power, higher MCS levels may be selected.
% Briefly mention the TPC mechanism from 802.11h
% Briefly mention the BSS color mechanism from 802.11ah

% Related work with SR in WiFi
% Related work can be moved to a separate section

\section{IEEE 802.11ax Protocol}
\label{80211ax}
In 2013, one of the subgroups of 802.11 group, named high-efficiency WLAN study group (HEW SG), began to study on a next-generation high-efficiency WLAN standard. Finally, in 2014, task group ax (TGax) was formed, and first draft of the 802.11ax standard has been published~\cite{Tgax}. 802.11ax protocol is the successor of 802.11ac, and the main goal is increasing the spatial efficiency and throughput per area in dense networks~\cite{80211ac}.

Starting with PHY layer changes and additions, WiFi6 offers a multitude of changes regarding the PHY layer. Although the primary goal of the IEEE 802.11ax protocol is spectral efficiency in dense environments, which contains overlapping BSSs, there are also two new MCS levels added to the standard. Accordingly, the data rate increases up to the gigabit level via 1024-QAM with a single antenna.

Initially, 802.11ax is planned to work only with classical WiFi channels in 2.4~Ghz and 5~GHz frequency bands, but with the extension called WiFi-6E, the new 6~GHz frequency band has been added recently~\cite{80211axStandard}. In addition to the frequency bands, 802.11ax has three guard interval options; \SI{0.8}{\micro\second}, \SI{1.6}{\micro\second} and \SI{3.2}{\micro\second}. Finally, IEEE 802.11ax supports Multi-User - Multiple Input Multiple Output (MU-MIMO) in both uplink (UL) and downlink (DL) directions, unlike 802.11ac, which has MU-MIMO only in DL direction. 

In addition to these PHY layer improvements, WiFi6 also has a pretty high number of new MAC layer mechanisms. Foremost of these mechanisms is the adoption of OFDMA. The primary rationale behind adding OFDMA support to WiFi is to allow simultaneous transmission in the DL and UL direction to reduce delay and increase overall efficiency. In OFDMA, multiple nodes can transmit simultaneously, reducing the interference by dividing the available resource units.

Another critical MAC layer addition of WiFi6 is the improved support for spatial reuse by BSS coloring that was introduced in 802.11ac and 802.11ah protocols before. According to the color mechanism of the 802.11ax protocol, each BSS has its color defined in the management frame with 6 bits of information. Since the BSS color header is 6 bits long, the protocol can assign 63 available colors to each BSS and its nodes in a distributed manner. The color information is carried by $\text{HE-SIG-A}$ frame that is contained by the HE-PPDU structure of 802.11ax. If there are overlapping BSSs, the color information of the transmitted frames can increase spectral efficiency by increasing the hearing threshold and ignoring the incoming packets from the Inter-BSS nodes. This threshold mechanism is called OBSS/PD.

\subsection{OBSS/PD Mechanism}
IEEE 802.11ax amendment focuses on spectral efficiency, and one of the critical features of the standard is the OBSS/PD mechanism. In dense environments, if there are overlapping BSSs that are using the same channels, the traffic belonging to these BSSs must share the channel and contend for access. Before performing the transmission, each STA should check whether the medium is in an idle state or not. In some cases, two STAs belonging to different BSSs cannot transmit simultaneously even if their transmissions will not cause transmission collisions. Hence, reducing the spectral efficiency.  

\begin{figure}[ht]
	\begin{center}
		\includegraphics[width=0.5\textwidth, keepaspectratio]{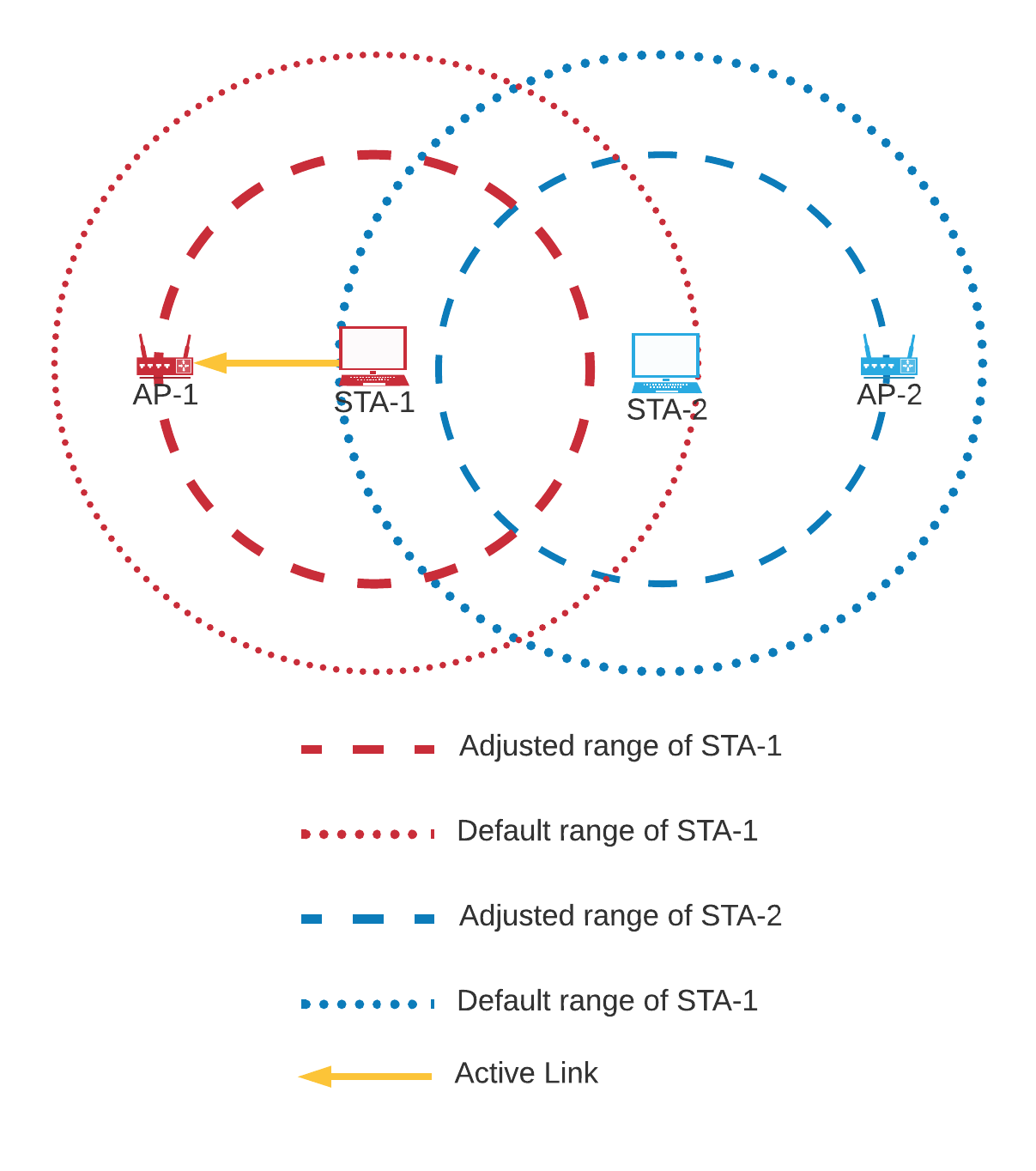}
		\caption{An example of spectral efficiency problem.}
		\label{fig:obsspd_threshold}
		\vskip\baselineskip
	\end{center}
\end{figure}

In Figure~\ref{fig:obsspd_threshold}, such a problematic scenario is illustrated when there are two different BSSs in close vicinity. The dotted circles represent the range in which a WiFi signal can be heard, red dotted circle for STA-1, and blue dotted circle for STA-2. Outside of these ranges, a WiFi transmission can not be heard. In this example, the active UL transmission between STA-1 and AP-1 causes interference to STA-2, and STA-2 cannot start communication with AP-2. Such concurrent transmissions will not have affected the reception process, so we are reducing the spectral efficiency. It is possible to overcome this problem if we had the option of adjusting each STA's ranges that affect which frames are received/heard. To continue this example, if the ranges are to be adjusted as the dashed bold circles, the STAs cannot receive the intended signal from each other, and both STAs can transmit data simultaneously.
%%%%%%%%%%%%%%%%%%%%%%%%%%%%%%%%%%%%%%%%%%%%%%%5

At its core, the spatial reuse features of 802.11ax allow the STAs and APs to adjust their ranges via transmit power and OBSS/PD thresholds dynamically to maximize spectral efficiency. OBSS/PD mechanism is an extension to the legacy CSMA/CA mechanism. 

\begin{figure}[ht]
	\begin{center}
		\includegraphics[width=0.45\textwidth, keepaspectratio]{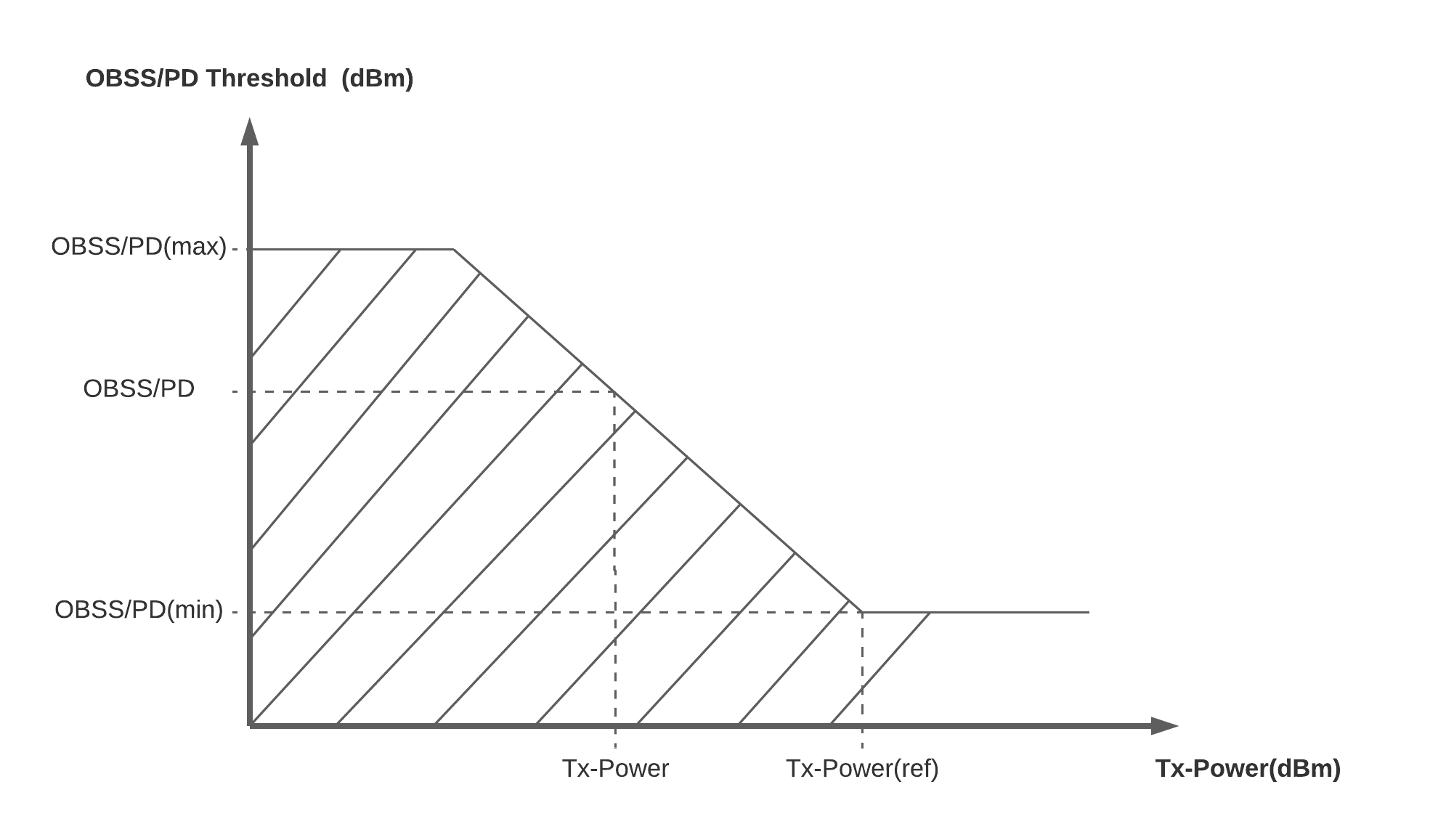}
		\caption{OBSS/PD-Transmision power relationship.}
		\label{fig:obsspd_graph}
		\vskip\baselineskip
	\end{center}
\end{figure}

The value of the OBSS/PD threshold level is directly related to Tx-Power. In~\cite{txobsspd}, it is shown that when the multiplication of transmission power and carrier sensing threshold is a constant value,  the wireless channel utilization/efficiency increases. In Figure~\ref{fig:obsspd_graph}, this relationship is illustrated for OBSS/PD threshold mechanism. If the Inter-BSS-based collision signal level is low, the OBSS/PD level should be set to lower levels, and at the same time, Tx-Power should be set to higher levels. On the other hand, if the medium has so many Inter-BSS collisions and the signal levels are high, in this case, the OBSS/PD should be set to higher levels to ignore these signals, and transmission power should be reduced, i.e., the communications should be carried over in close vicinity with a lower transmit power and ignoring interference signals. 
%%%%%%%%%%%%%%%%%%%%%%%%%%%%%%%%%%%%%%%%%%%%%%%%

From the relationship of OBSS/PD threshold and transmission power, the transmission power is calculated via the given OBSS/PD level according to the Equation~\ref{eq:obss_eq}. Since there is a linear relationship between OBSS/PD threshold and transmission power, as the transmission power increases, OBSS/PD should be decreased the same amount of value or vice versa. For each channel bandwidth (i.e., \SI{20}{\mega\hertz}, \SI{40}{\mega\hertz}, \SI{80}{\mega\hertz}, and \SI{160}{\mega\hertz}), there are upper and lower bounds for OBSS/PD threshold and transmission power levels. In Table~\ref{table:obsspd_range}, the minimum and maximum limits of OBSS/PD threshold and transmission power ($TxPow$) are given for the corresponding channel bandwidths. These limits determine the region that an algorithm can adjust the OBSS/PD threshold and $TxPow$ accordingly.
\begin{align}
    \label{eq:obss_eq}
    OBSS/PD &\!=\! OBSS/PD_{Min} \!+\! TxPow_{Ref} \!-\! TxPow
\end{align}
where $TxPow_{Ref}$ is the maximum transmission power and $OBSS/PD_{Min}$ is the minimum OBSS/PD threshold that is allowed to be set.
%%%%%%%%%%%%%%%%%%%%%%%%%%%%%%%%%%%%%%%%%%%%%%%%%
\begin{table}[t]
%\vskip\baselineskip 
\caption[OBSS/PD ranges in \si{\dBm} at different channel bandwidths.]{OBSS/PD ranges in \si{\dBm} at different channel bandwidths.}
\begin{center}
	\begin{tabular}{|c|c|c|c|c|}\hline
		\textbf{}& \textbf{\SI{20}{\mega\hertz}}& \textbf{\SI{40}{\mega\hertz}}& \textbf{\SI{80}{\mega\hertz}}&\textbf{\SI{160}{\mega\hertz}}\\\hline
		$OBSS/PD_{Min}$ & -82.0 & -79.0 & -76.0 & -73.0\\\hline
		$OBSS/PD_{Max}$ & -62.0 & -59.0 & -56.0 & -53.0\\\hline
		$TxPow_{Ref}$ & 21.0 & 18.0 & 15.0 & 12.0\\\hline
	\end{tabular}
\label{table:obsspd_range}
\end{center}
\end{table}
%%%%%%%%%%%%%%%%%%%%%%%%%%%%%%%%%%%%%%%%%%%%%%%%%

\begin{figure}[ht]
	\begin{center}
		\includegraphics[width=0.5\textwidth, keepaspectratio]{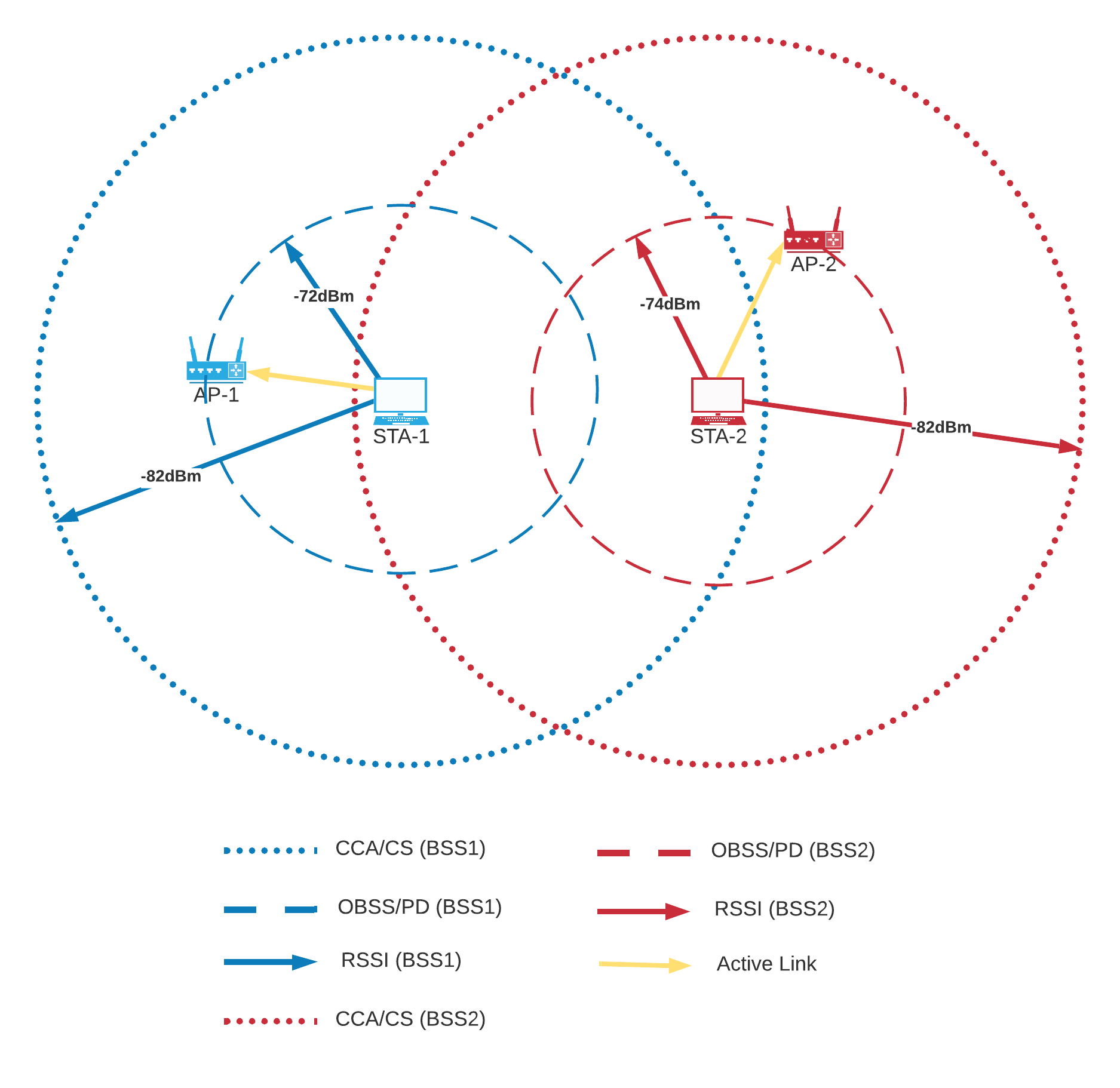}
		\caption{A simple scenario for OBSS/PD mechanism.}
		\label{fig:obsspd_scenario}
		\vskip\baselineskip
	\end{center}
\end{figure}

In Figure~\ref{fig:obsspd_scenario}, there are two BSSs, and each of them has one AP. Each station is associated with an AP and performs an uplink transmission. The default CCA Carrier Sensing thresholds are set to \SI{-82}{\dBm}, and both transmissions interfere with each other. This issue creates a significant amount of performance degradation in the overall system. In this problematic case, since the interfering signals are originated from Inter-BSS stations, OBSS/PD mechanism can be used to eliminate this interference. Suppose a node increases both OBSS/PD thresholds to \SI{-72}{\dBm} and \SI{-74}{\dBm}, respectively. In that case, it will ignore the incoming Inter-BSS signal under this level, and it can initiate transmission with their corresponding APs. According to the conditions and system parameters, while keeping the communication link alive, OBSS/PD level can be increased, and the same amount of transmission power can be decreased to reduce the interference. That means the incoming RSSIs from the Inter-BSS stations will be reduced at the receiver's physical layer, and the incoming signals will not collide because they are under the OBSS/PD threshold.
%%%%%%%%%%%%%%%%%%%%%%%%%%%%%%%%%%%%%%%%%%%%%%%%%%%

%%%%%%%%%%%%%%%%%%%%%%%%%%%%%%%%%%%%%%%%%%%%%%%%%%%%%%%%
%%%%%%%%%%%%%%%%%%%%%%%%%%%%%%%%%%%%%%%%%%%%%%%%%%%%%%%%
%%%%%%%%%%%%%%%%%%%%%%%%%%%%%%%%%%%%%%%%%%%%%%%%%%%%%%%%
\section{RACEBOT ALGORITHM}
\label{chapter:Racebot}
In this section, our proposed algorithm, RACEBOT, will be explained. RACEBOT algorithm is an OBSS/PD threshold-based algorithm, which dynamically adjusts OBSS/PD threshold and the transmit power in the run-time. The main goal of the RACEBOT algorithm is to find an effective OBSS/PD threshold and consequently transmit power to increase aggregated throughput while reducing Inter-BSS interference and reducing the possibility of hidden STAs.

Adjusting the OBSS/PD threshold considering RSSI can be implemented in various ways. As discussed earlier, some of these algorithms are based on beacon RSSI, and some of the other algorithms use both beacon RSSI and RSSI of OBSS STAs. RACEBOT algorithm also uses both RSSI types for two reasons: using beacon RSSI to keep the link with the associated AP alive and using OBSS RSSI to avoid OBSS interference. RACEBOT algorithm records each RSSI with their OBSS frame count ($OFC$) in a specific period. These $OFC$ values are used to adjust OBSS/PD thresholds in the following stages of the algorithm. 

To select the appropriate OBSS/PD threshold, RACEBOT algorithm sets an $OFC$ threshold named $OFC_{Thr}$ and determines the maximum RSSI, whose count is greater than the $OFC_{Thr}$ value. 

RACEBOT algorithm consists of two parts: finding a goal OBSS/PD threshold and adjusting the threshold by increasing the OBSS/PD value step by step with considering the effects of the change in transmission power. Figure~\ref{alg:racebot} shows the working mechanism of the RACEBOT algorithm. First of all, $M$, $\gamma$, $a$, $T_{U}$, $OFC_{Thr}$, $RSSI^{OBSS}$, $\overline{RSSI}^{BSS}$, $OBSS/PD^{G}$,  $\overline{MCS}$ and $\overline{OFC}$ parameters are initialized. $T_{U}$ is a periodic time interval for the algorithm to update some of the run-time parameters. If the timer $T$ is greater than $T_{U}$, $T$ is reset and goal OBSS/PD ($OBSS/PD^{G}$) value of each node is initialized with -101 dBm. The $OBSS/PD^{G}$ parameter is the aimed OBSS/PD threshold that the RACEBOT algorithm tries to reach in run-time. For each RSSI, EWMA of $OFC$ values are defined by $\overline{OFC}$ parameter as in the Equation~\ref{eq:ewma_conflict}. 
%%%%%%%%%%%%%%%%%%%%%%%%%%%%%%%%%%%%%%%%%%%%%%%%%
\begin{equation}
\label{eq:ewma_conflict}
  \overline{OFC}(T) = \alpha \times OFC(T) + (1-\alpha) \times \overline{OFC}(T - 1)
\end{equation}
%%%%%%%%%%%%%%%%%%%%%%%%%%%%%%%%%%%%%%%%%%%%%%%%%%%%%%%%%%%%%%%%%%%%%%%%%%%%%%%%%%%%%%%%%%
\begin{algorithm*}
\begin{center}
\begin{algorithmic}[1]
\STATE \textbf{Initialize} $M$, $\gamma$, $\alpha$, $T_{U}$, $OFC_{Thr}$, $RSSI^{OBSS}$, $\overline{RSSI}^{BSS}$, $OBSS/PD^{G}$,  $\overline{MCS}$, $\overline{OFC}$
\WHILE{RUN}
\IF{$T>T_{U}$}
\STATE \textbf{Set} $T=0$ , $OBSS/PD^{G}_{i}=-101dBm$ where $i=1,2.....n$
\FOR{ Each $Node_{i}$  }
\FOR{ Each $RSSI^{OBSS}_{i,j}$ , $j=1,2....m$ }
\IF{ $\overline{OFC}_{i,j}(T)>OFC_{Thr}$\quad and\quad $RSSI^{OBSS}_{i,j}>OBSS/PD^{G}_{i}$}
\STATE $OBSS/PD^{G}_{i}$=min($RSSI^{OBSS}_{i,j}+M$, $\overline{RSSI}^{BSS}_{i}-M$)
\ENDIF
\ENDFOR
\ENDFOR
\STATE \textbf{Reset} $RSSI^{OBSS}$, $\overline{OFC}$
\ENDIF
\STATE
\STATE \textbf{Collect} $\overline{RSSI}^{BSS}$ and $RSSI^{OBSS}$
\IF{$\overline{MCS}(T-1)\times \gamma > \overline{MCS}(T)$}
\STATE $OBSS/PD= max(\frac{(OBSS/PD +\overline{RSSI}^{BSS}-M)}{2},OBSS/PD_{min})$
\STATE $OBSS/PD^{G}= \frac{(OBSS/PD +OBSS/PD^{G})}{2}$
\ELSIF{$\overline{MCS}(T-1)\times \gamma <= \overline{MCS}(T)$}
\STATE $OBSS/PD= min(\frac{(OBSS/PD + OBSS/PD^{G})}{2},OBSS/PD_{max})$
\ENDIF
\ENDWHILE
\end{algorithmic}
\end{center}
\caption[RACEBOT algorithm.]{RACEBOT algorithm.}
% \vskip
\label{alg:racebot}
\end{algorithm*}

To calculate the $OBSS/PD^{G}$ parameter, first the $OFC$ values for each RSSI are compared to $OFC_{Thr}$ parameter, which is a predefined positive real number from $\mathbb{R}^{+}$. $OFC_{Thr}$ value helps to prune the received RSSI values with considering the rareness by $OFC$ values. $\overline{RSSI}^{BSS}$ is the EWMA of the beacon RSSI of associated AP. RACEBOT also uses $M$ value, which is a positive real number from $\mathbb{R}^{+}$ to be used as  a margin. For a single node, the algorithm looks whether  $OFC$ value of corresponding RSSI is greater than $OFC_{Thr}$ and $RSSI^{OBSS}$ is bigger than $OBSS/PD^{G}$ or not. If so, the $M$ value is added to $RSSI^{OBSS}$ and subtracted from $\overline{RSSI}^{BSS}$ then minimum of them is assigned to the $OBSS/PD^{G}$ parameter. It is repeated for all the nodes in the system. Then $RSSI^{OBSS}$ and $\overline{OFC}$ are reset to collect new ones. RACEBOT algorithm follows the changes of MCS level. 

As discussed earlier, each MCS level requires a minimum RSSI and maximum PER at the receiver side. Therefore, increasing the OBSS/PD threshold reduces the transmission power, and it may prevent using higher MCS levels. To overcome this issue, RACEBOT algorithm collects EWMA of MCS levels and compares the current average MCS level ($\overline{MCS}(T)$) with the average MCS level of the previous step time ($\overline{MCS}(T-1)$). In this step, there is also a constant $\gamma$, which is a real number between 0 and 1. The $\gamma$ parameter is multiplied by the $\overline{MCS}(T-1)$ to compare how much change happened to $\overline{MCS}$. For example, if $\gamma$ is $0.7$, it is checked that whether the $\overline{MCS}$ value decreases or not by more than 30\%. 

Algorithm~\ref{alg:racebot} shows the adjusting mechanism of the RACEBOT algorithm for the OBSS/PD threshold. If $\overline{MCS}$ value decreases for a given criterion, OBSS/PD threshold is reduced. In order to reduce OBSS/PD level, an average of the current OBSS/PD threshold or $\overline{RSSI}^{BSS}$ are assigned as a new OBSS/PD threshold. If this value is above the $OBSS/PD_{min}$ level, then $OBSS/PD^{G}$ is reduced to the average of the new OBSS/PD threshold and $OBSS/PD^{G}$. If $\overline{MCS}$ does not decrease for a given criterion, OBSS/PD threshold is increased and the average of the OBSS/PD threshold. $OBSS/PD^{G}$ value is assigned as OBSS/PD threshold, if it is less than $OBSS/PD_{max}$. In Algorithm~\ref{alg:racebot}, the $OBSS/PD_{min}$ and $OBSS/PD_{max}$ boundary values are used in the comparisons at lines 18 and 20 to prevent any inconsistencies in the signal power in case of the existence of a signal that is out of the OBSS/PD boundary.

The main contributions of the RACEBOT algorithm can be summarized as follows:

\begin{enumerate}
  \item RACEBOT algorithm is adaptive to selected data rates by rate selection algorithms, and it creates a chance for the upper MCS levels that rate algorithms can use. Therefore transmission can be performed via higher throughput.
  \item The algorithm decreases the number of exposed nodes by adjusting an effective OBSS/PD threshold.
  \item RACEBOT algorithm does not decrease the transmission power sharply. Therefore, reducing the number of exposed nodes keeps the number of hidden nodes at acceptable levels.
  \item The algorithm can work with other rate selection algorithms.
  \item The Algorithm works at both sparse and high-density networks without having a problem.
\end{enumerate}

\section{System Parameters and Performance Metrics}
In the simulations, the parameters used in the RACEBOT algorithm are as given in Table~\ref{table:parameters}. As discussed earlier, OBSS/PD mechanism is proposed in the IEEE 802.11ax amendment. Since the RACEBOT algorithm is based on OBSS/PD mechanism, 802.11ax protocol has been used in the simulations. Among the three operational frequencies in 802.11ax, 2.4 GHz, 5 GHz, and 6 GHz, we choose the 5 GHz frequency band. As the modulation schema, OFDM is used with 20 MHz channel bandwidth. As a guard interval (GI), \SI{0.8}{\micro\second} short GI is used. As for spatial streams, all devices are set to have a single spatial stream (i.e., a 1x1 MIMO scheme).

The direction of the traffic is set to be the UL direction. A constant bit rate (CBR) traffic combined with a Poisson traffic model is used as the traffic generator. Since delay is not a concern of this study, 300 Mbps constant data rate and 1024 bytes payload size are used to saturate the traffic buffer. Since the RACEBOT algorithm is designed to be used with rate selection algorithms, in the scenario, a modified version of Minstrel-HT for 802.11ax HE-rates and Thompson Rate Algorithm that are inside ns3-dev are used as rate selection algorithms. In the next chapter, one of the TGax outdoor scenarios is used. Also, based on this scenario, additional random custom scenarios are generated and used. As OBSS/PD threshold boundaries, In 20 MHz bandwidth, default $OBSS/PD_{min}$ and $OBSS/PD_{max}$ values -82 dBm and -62 dBm are used.

\begin{table}[thbp]
%\vskip\baselineskip 
\caption[System parameters.]{System parameters.}
\begin{center}
    \begin{tabular}[c]{|*{2}{c|}}   % repeats {c|} 18 times
        \hline 
        \multicolumn{1}{|c}{\textbf{Parameter}} &\multicolumn{1}{|c|}{\textbf{Value}}\\\hline
        \textbf{Protocol Standard} & 802.11ax \\\hline
        \textbf{Frequency} & 5 GHz \\\hline
        \textbf{Modulation Schema} & OFDM \\\hline
        \textbf{Bandwidth} & 20 MHz \\\hline
        \textbf{Guard Interval} & Short (\SI{0.8}{\micro\second}) \\\hline
        \textbf{Traffic Type} & UDP CBR Poisson \\\hline
        \textbf{Traffic Direction} & Uplink \\\hline
        \textbf{Rate} & 300 Mbps \\\hline
        \textbf{Payload Size} & 1024 Bytes \\\hline
        \textbf{MCS selection} & (Minstrel, Thompson) \\\hline
        \textbf{Scenario} & Original Box5 and Custom Box5 \\\hline
        \textbf{Number of Antennas} & 1 \\\hline
        \textbf{$OBSS/PD_{min}$,$OBSS/PD_{max}$} & -82 dBm, -62 dBbm \\\hline
        \textbf{Preamble Detection Threshold} & -82 dBm \\\hline
        \textbf{RxSensitivity} & -82 dBm\\\hline
        \textbf{RTS/CTS} & Disabled \\\hline
        \textbf{CCA/ED} & -62 dBm \\\hline
        \textbf{TxPower AP/STA} & 21 dBm / 21 dBm \\\hline
        \textbf{Simulation Time} & 50 s \\\hline
        \textbf{Loss Model} & Friis Loss Model \\\hline
        \textbf{OFC} & 10 \\\hline
        \textbf{$\gamma$} & 0.7 \\\hline
\end{tabular}
\label{table:parameters}
\end{center}
\end{table}

The default preamble detection and $RxSensitivity$ thresholds that NS-3 uses are adjusted to -82 dBm, and the energy detection threshold $CCA/ED$ is adjusted to -62 dBm. The initial $TxPower$ of both AP and STA's are set to 21 dBm, but at the run-time, each $TxPower$ of nodes may change according to variations in the OBSS/PD thresholds due to our algorithm. As the loss model, we used Friis free space propagation model. This model works for the far-field region. Because of this, the scenarios are designed to operate according to this condition. Finally, all the simulations run for 50 seconds simulation time.

In the simulations, our performance metric is aggregated throughput, which is defined as the overall throughput of the system. Throughput of a single node in Mbits is calculated as 
\begin{equation}
\label{eq:throughput}
  Throughput_{i}= \frac{rBytes_{i} \times 8}{1000 \times 1000 \times T_{Step}} \quad where\quad i=0,1,2...,
\end{equation}
where $rBytes_{i}$ is the total received data in bytes for each $T_{Step}$ and $T_{Step}$ is the step time. Then, the aggregated throughput of all nodes in the system for the given time step is evaluated as
%%%%%%%%%%%%%%%%%%%%%%%%%%%%%%%%%%%%%%%%%%%%%%%%%%%
\begin{equation}
\label{eq:aggthroughput}
  Throughput^{Agg}_{s} = \sum_{i=1}^{\\n} Throughput_{i},
\end{equation}
%%%%%%%%%%%%%%%%%%%%%%%%%%%%%%%%%%%%%%%%%%%%%%%%%%%
where $n$ is the number of nodes in the system. Finally, the aggregated total data that is transferred in the total simulation time is calculated as
%%%%%%%%%%%%%%%%%%%%%%%%%%%%%%%%%%%%%%%%%%%%%%%%%%%
\begin{equation}
\label{eq:tottransmission}
  TM_{Tot} = T_{Step} \times \sum_{s=0}^{\\ N_{Step}}Throughput^{Agg}_{s}
\end{equation}
%%%%%%%%%%%%%%%%%%%%%%%%%%%%%%%%%%%%%%%%%%%%%%%%%%%
where $N_{Step}$ is defined as the total number of steps in the whole simulation.

\section{Simulation Topology}
\label{topology}
In this section, the topology that is used in the simulations will be discussed. In~\cite{tgaxbox5}, there are multiple scenarios developed by the collaboration of different product vendors inside TGax to evaluate and calibrate the performance of the systems. 
%%%%%%%%%%%%%%%%%%%%%%%%%%%%%%%%%%%%%%%%%%%%%%%%%%%
\begin{figure}[htbp]
	\begin{center}
		\includegraphics[width=0.5\textwidth]{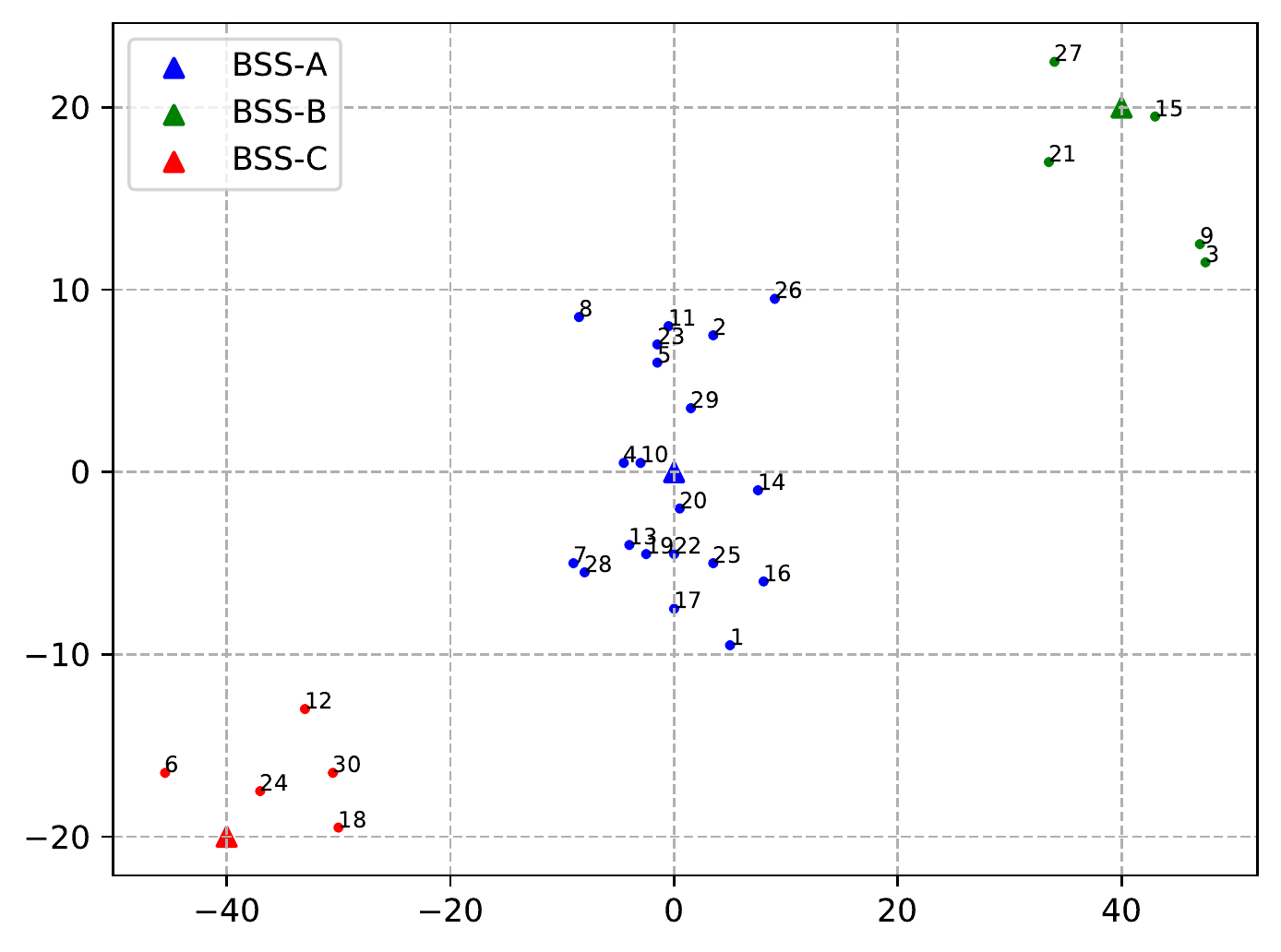}
		\vskip\baselineskip % Leave a vertical skip below the figure
		\caption{TGax Box5 outdoor topology.}
		\label{fig:box5}
	\end{center}
\end{figure}
%%%%%%%%%%%%%%%%%%%%%%%%%%%%%%%%%%%%%%%%%%%%%%%%%%%

In Figure~\ref{fig:box5}, you can see one of the sophisticated outdoor scenarios in the TGax test suite named ``Box5" is shown~\cite{tgaxbox5}. This topology consists of three BSSs (i.e., 3 APs) and 30 STA. While BSS-A has 20 associated STAs, BSS-B and BSS-C have 5 STAs each. In this study, the simulations are performed based on this topology.

For the ``TGax Box5" scenario, a single topology inside the TGax documentation is used~\cite{tgaxbox5}. Since this topology is used for performance comparisons and calibration, the corresponding result of the simulations may give some inferences. 

\section{Simulation Results}
We considered different scenarios to evaluate the performance of the RACEBOT algorithm. Real-life applications in wireless networks take advantage of rate selection algorithms to get maximum performance from the system. In the scenarios of the thesis study, a modified version of the Minstrel-HT algorithm for HE-rates and Thompson rate selection algorithm with $decay=0.1$ are used as reference algorithms\cite{Minstrel-HT}\cite{Thompson}. Since each rate algorithm has a different performance, the simulations are evaluated for each rate selection algorithm separately.

As reference algorithms, we used the DSC algorithm, RTOT algorithm, and a scenario (named NO-OBSSPD) that does not use any carrier sensing threshold mechanism to compare the RACEBOT algorithm with the other carrier sensing threshold algorithms in the scenarios\cite{threshold6}\cite{threshold13}.

\subsection{Box5 Scenario}
TGax defined 'Box5 Scenario' as an outdoor scenario, as mentioned at the beginning of Section~\ref{topology}.  'Box5 Scenario' consists of three BSSs as BSS-A contains 1 AP and 15 STAs, each of BSS-B and BSS-C has 1 AP and 5 STAs (Figure~\ref{fig:box5}).

\begin{figure}[htbp]
	\begin{center}
		\includegraphics[width=0.5\textwidth, keepaspectratio]{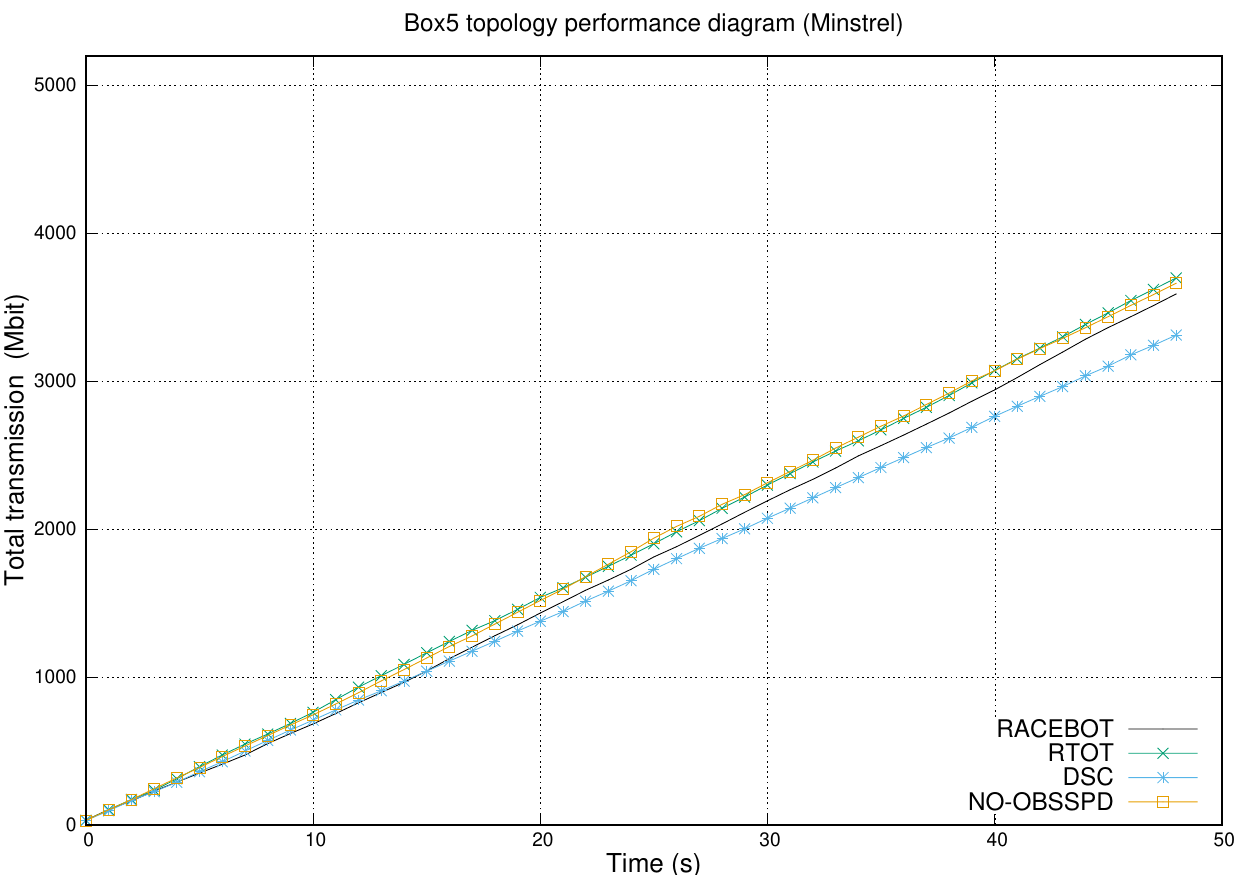}
		\caption{Scenario Box5 - Minstrel.}
		\label{fig:ScenarioBox5Minstrel}
		\vskip\baselineskip
	\end{center}
\end{figure}

Figure~\ref{fig:ScenarioBox5Minstrel} illustrates the simulation results of RACEBOT, RTOT, DSC, and NO-OBSSPD with Minstrel rate selection algorithm. Here, the DSC algorithm gives a reduced total throughput compared to the other algorithms. As for our RACEBOT algorithm, it starts with slightly worse performance than RTOT and NO-OBSSPD algorithms, but after \SI{42}{\second}, the slope of the curve increases, and its results converge to that of RTOT and NO-OBSSPD.
%%%%%%%%%%%%%%%%%%%%%%%%%%%%%%%%%%%%%%%%%%%%%%%%%%%%%%

\begin{figure}[htbp]
	\begin{center}
		\includegraphics[width=0.5\textwidth, keepaspectratio]{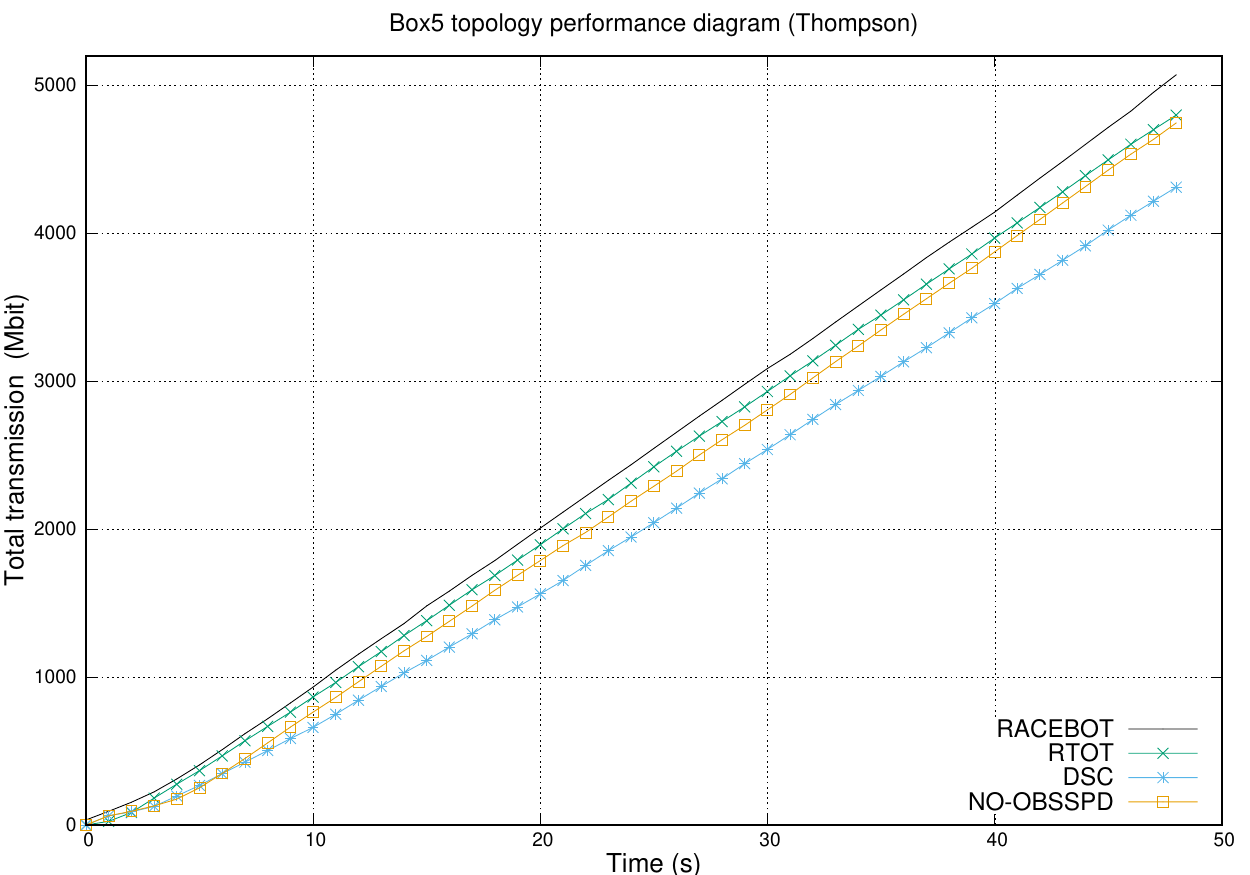}
		\caption{Scenario Box5 - Thompson.}
		\label{fig:ScenarioBox5Thompson}
		\vskip\baselineskip
	\end{center}
\end{figure}

When we switch to the Thompson rate selection algorithm in the same scenario, the results change considerably, as seen in Figure~\ref{fig:ScenarioBox5Thompson}. Again, the DSC algorithm has shown worse performance than the other algorithms, and again, the RTOT algorithm and NO-OBSSPD case have similar performance. However, our proposed algorithm, RACEBOT, has shown a slightly better result than the other three algorithms, and the discrepancy keeps increasing as time passes. It indicates that the DSC algorithm has some limitations in both rate selection algorithms. It even shows worse performance than the case that does not use any carrier sensing threshold algorithm. 
%%%%%%%%%%%%%%%%%%%%%%%%%%%%%%%%%%%%%%%%%%%%%%%%

%%%%%%%%%%%%%%%%%%%%%%%%%%%%%%%%%%%%%%%%%%%%%%%%%%%%%%%%
%%%%%%%%%%%%%%%%%%%%%%%%%%%%%%%%%%%%%%%%%%%%%%%%%%%%%%%%
%%%%%%%%%%%%%%%%%%%%%%%%%%%%%%%%%%%%%%%%%%%%%%%%%%%%%%%%
\section{Contributions of RACEBOT Algorithm}
In this section, the contributions of the RACEBOT algorithm and the potential future directions that can be improved based on these results are discussed.

RACEBOT algorithm uses the advantage of controlling transmission power and having information of conflicts to increase the overall performance. Since the transmission power is directly related to the data rate, the RACEBOT algorithm does not decrease the transmission power sharply. It creates a chance for the rate selection algorithm to use higher rates. 

RACEBOT algorithm uses $OFC$ and $OFC_{Thr}$ parameters for pruning exceptional RSSI values. These parameters are used to neglect RSSIs, which are rare and can degrade the performance. In that way, the RACEBOT algorithm can use higher transmission powers and higher data rates. 

Sharp changes in transmission power may degrade the performance of rate selection algorithms. Therefore, the RACEBOT algorithm does not increase the OBSS/PD level sharply; it increases the OBSS/PD level step by step by considering the changes of MCS levels. RACEBOT algorithm keeps track of MCS deterioration while changing the OBSS/PD level. 

Our algorithm decreases the number of exposed nodes by lowering the OBSS/PD threshold. Since the RACEBOT algorithm varies the OBSS/PD level step by step, it can keep the hidden nodes lower than the algorithms changing the OBSS/PD level sharply.

%%%%%%%%%%%%%%%%%%%%%%%%%%%%%%%%%%%%%%%%%%%%%%%%%%%%%%%%
%%%%%%%%%%%%%%%%%%%%%%%%%%%%%%%%%%%%%%%%%%%%%%%%%%%%%%%%
%%%%%%%%%%%%%%%%%%%%%%%%%%%%%%%%%%%%%%%%%%%%%%%%%%%%%%%%
\section{Future Directions}
In a coordinated infrastructure, as a future direction, this mechanism can be modified and coordinated by a central location. The power, location, and collision information of each node can be used to determine an optimal OBSS/PD threshold. We plan to establish a system that can be coordinated centrally to operate OBSS/PD mechanism

Besides this, we plan to develop a new rate selection algorithm that can use this OBSS/PD mechanism internally. In that way, nodes will use the OBSS/PD mechanism with another rate selection algorithm or inside the new rate selection algorithm by choice.

We plan to optimize the parameters of the RACEBOT algorithm for different situations, and we will try to design an auto-adjustment algorithm for self-adaptation. Moreover, we will design a new algorithm with adaptive step size for varying the OBSS/PD level while moving towards the goal-OBSS/PD. 
\ifCLASSOPTIONcaptionsoff
  \newpage
\fi



%\bibliographystyle{IEEEtran}
\bibliographystyle{ieeetr}
\bibliography{references}

\begin{thebibliography}{1}

\bibitem{Tgax}
{Task Group-ax}, ``High efficiency (he) wireless lan task group.''
  \url{https://www.ieee802.org/11/Reports/tgax_update.htm}, 2014.

\bibitem{80211ac}
{IEEE 802.11ac Task Group}, ``Ieee standard for information technology--
  telecommunications and information exchange between systemslocal and
  metropolitan area networks-- specific requirements--part 11: Wireless lan
  medium access control (mac) and physical layer (phy)
  specifications--amendment 4: Enhancements for very high throughput for
  operation in bands below 6 ghz.,'' {\em IEEE Std 802.11ac-2013 (Amendment to
  IEEE Std 802.11-2012, as amended by IEEE Std 802.11ae-2012, IEEE Std
  802.11aa-2012, and IEEE Std 802.11ad-2012)}, pp.~1--425, 2013.

\bibitem{80211axStandard}
{IEEE 802.11ax Task Group}, ``Ieee standard for information
  technology--telecommunications and information exchange between systems local
  and metropolitan area networks--specific requirements part 11: Wireless lan
  medium access control (mac) and physical layer (phy) specifications amendment
  1: Enhancements for high-efficiency wlan,'' {\em IEEE Std 802.11ax-2021
  (Amendment to IEEE Std 802.11-2020)}, pp.~1--767, 2021.

\bibitem{txobsspd}
J.~A. Fuemmeler, N.~H. Vaidya, and V.~V. Veeravalli, ``Selecting transmit
  powers and carrier sense thresholds in csma protocols for wireless ad hoc
  networks,'' in {\em Proceedings of the International Workshop on Wireless
  Internet (WICON)}, (New York, NY, USA), p.~15–es, Association for Computing
  Machinery, 2006.

\bibitem{tgaxbox5}
S.~Merlin, ``Tgax simulation scenarios.''
  \url{https://mentor.ieee.org/802.11/dcn/14/11-14-0980-16-00ax-simulation-scenarios.docx},
  2015.
\newblock {IEEE 802.11ax , IEEE 802.11-14/0980r1}.

\bibitem{Minstrel-HT}
F.~Fietkau, ``Minstrel-ht rate algorithm.''
\newblock {Accessed in July 2021}.

\bibitem{Thompson}
H.~Gupta, A.~Eryilmaz, and R.~Srikant, ``Low-complexity, low-regret link rate
  selection in rapidly-varying wireless channels,'' in {\em Proceedings of the
  IEEE International Conference on Computer Communications (INFOCOM)},
  pp.~540--548, 2018.

\bibitem{threshold6}
G.~Smith, ``Dynamic sensitivity control-v2.''
  \url{https://mentor.ieee.org/802.11/dcn/13/11-13-1012-04-0wng-dynamic-sensitivity-control.pptx},
  2013.
\newblock {IEEE 802.11-13/1012r4}.

\bibitem{threshold13}
T.~Ropitault, ``Evaluation of rtot algorithm: A first implementation of obss
  pd-based sr method for ieee 802.11ax,'' in {\em Proceedings of the IEEE
  Annual Consumer Communications Networking Conference (CCNC)}, pp.~1--7, 2018.

\end{thebibliography}


%\vfill


\end{document}